# Gaze-Driven Adaptive Interventions for Magazine-Style Narrative Visualizations

Sébastien Lallé, Dereck Toker, and Cristina Conati

**Abstract**—In this paper we investigate the value of gaze-driven adaptive interventions to support processing of textual documents with embedded visualizations, i.e., Magazine Style Narrative Visualizations (MSNVs). These interventions are provided dynamically by highlighting relevant data points in the visualization when the user reads related sentences in the MNSV text, as detected by an eye-tracker. We conducted a user study during which participants read a set of MSNVs with our interventions, and compared their performance and experience with participants who received no interventions. Our work extends previous findings by showing that dynamic, gaze-driven interventions can be delivered based on reading behaviors in MSNVs, a widespread form of documents that have never been considered for gaze-driven adaptation so far. Next, we found that the interventions significantly improved the performance of users with low levels of visualization literacy, i.e., those users who need help the most due to their lower ability to process and understand data visualizations. However, high literacy users were not impacted by the interventions, providing initial evidence that gaze-driven interventions can be further improved by personalizing them to the levels of visualization literacy of their users.

──────────♦──────────

## 1 INTRODUCTION

Visualizations are typically designed following a one size-fits-all approach, meaning that they do not take into account individual differences in their users. There is however mounting evidence that user characteristics such as cognitive abilities and personality traits, can significantly influence user experience during information visualization (InfoVis) tasks, even with well-designed, thoroughly evaluated visualizations, e.g., [1]–[4]. These findings have prompted researchers to study *user-adaptive information visualizations*, i.e., visualizations that can recognize specific needs and abilities of their individual users, and adapt various aspects of the visualization accordingly.

The first examples of adaptive visualizations leveraged user actions with an interactive visualization system to detect evidence that the user is not working well with the current visualization, and suggest a suitable alternative (e.g., [5]–[7]). More recently, researchers have been investigating eye-tracking data as a source of information to predict user needs and drive adaptation. Eye-tracking is especially suitable for delivering adaptation in visualization because it can directly capture visual processes that are fundamental for working with a visualization. Eye-tracking data has also the advantage of being available in both interactive and non-interactive visualizations.

Existing research has shown that eye-tracking data can be used to predict in real time several *long-term user characteristics* and *short-term states* known to influence visualization effectiveness, such as users' perceptual abilities [8]–[10], interest [11], confusion [12] and cognitive load [13]. Whereas these predictions of user long-term characteristic and short-term states have yet to be used for adaptation, there is initial evidence on the effectiveness of adaptation simply based on detecting specific user gaze behaviors, i.e., *gaze-driven* adaptation. Specifically, Göbel et al. [14] and Bektas et al. [15] have shown that *gaze-driven* adaptation can facilitate processing of map-based visualizations.

We contribute to this body of research by investigating gaze-driven adaptation as a means to support processing of visualizations embedded in narrative text, known as Magazine-Style Narrative Visualization (MSNV for short) [16]. We focus on MSNVs featuring bar charts, one of the most ubiquitous visualizations found in MSNV documents such as newspapers, scientific articles, blogs, textbooks [17]. We also investigate the potential value of long-term user characteristics to further personalize the delivery of gaze-drive adaptation in MSNVs.

As it is often the case for multimodal documents, processing MSNVs can be challenging due to the need to split attention between two information sources, with a possible increase in cognitive load and negative impact on comprehension [18]. The challenge can be exacerbated in MSNVs as there are often multiple sentences in the text, called *references*, that solicit attention to different aspects of the accompanying visualization (see example in Fig. 1). In particular, identifying which data points in the visualizations correspond to each reference is a well-known difficulty in MSMVs [18], [20]–[23].

Based on these results, in this paper we design and evaluate a form of cuing (i.e. adding visual prompts that guide user attention) that aims to facilitate MSNV processing by highlighting the relevant data points in the visualization

• *S. Lallé, D. Toker, and C. Conati are with the Department of Computer Science at the University of British Columbia, 2366 Main Mall, Vancouver, BC V6T 1Z4, Canada. E-mail: {lalles, dtoker, conati}@cs.ubc.ca.*

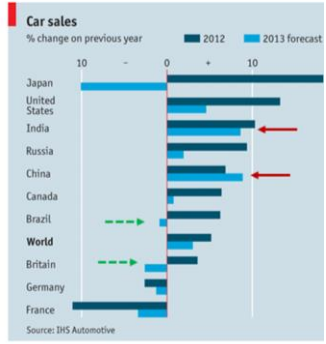

Fig. 1. An example of MSNV document with multiple references, with the first two underlined by us for easier identification. Arrows identify the different data points the underlined sentences refer to in the MNSV bar graph. Source: The Economist - Dec 22, 2012.

when a user reads a reference in the text, as detected by an eye-tracking device. We also investigate whether the effectiveness of this gaze-driven adaptation depends on user characteristics previously shown to exacerbate difficulties in processing bar-chart-based MSNV [23], to ascertain if the gaze-driven adaptation should be further personalized to any of these user characteristics. Thus, the research question we investigate in this paper is as follows:

*Does gaze-driven highlighting of relevant parts of an MSNV graph improve user performance and subjective experience with bar-chart-based MSNVs, compared to users who received no such highlighting? Do the results depend on user characteristics that were previously found to influence MSNV processing?*

To answer this question, we conducted a user study during which participants read a set of 14 MSNV with our proposed gaze-driven highlighting interventions, and we compared their outcomes with those of users who underwent the exact same task without interventions in the study described in [23].

Our results show that the proposed highlighting interventions specifically helped users with low levels of *visualization literacy (*vis literacy for short*)*, i.e., users with lower ability to process and understand data visualizations [24]. This finding is significant because it shows that the adaptive interventions benefited those users who needed help the most. Our results also show that users with high vis literacy were not impacted by the interventions, but their performance still had room for improvement. This suggests to further investigate adaptive interventions personalized to be helpful for these high vis literacy users.

Our work contributes to existing research in two ways. First, it broadens the evidence on the value of eye-tracking for adaptive visualizations. Prior to our work, gaze-driven adaptation has only been used to support the processing of map-based visualizations [14], [15]. Our results extend these findings both by looking at a different visualization type, bar charts, as well as by focusing on MSNV, a widespread context of usage for visualizations that has not yet been investigated for gaze-driven adaptation.

A second contribution is that our results are the first to show that *long-term user characteristics* can influence the effectiveness of gaze-driven adaptation in visualization, calling for further research on how these interventions can be personalized to these long-term characteristics.

## 2 RELATED WORK

### 2.1 Need for Adaptation in InfoVis

*Short-term states* such as *cognitive overload* [25] and *confusion* [26] can directly reveal when the user is struggling when performing a task with a visualization, signalling that real-time support addressing the specific user difficulties would be beneficial.

There is also extensive evidence that *long-term user characteristics* (e.g., cognitive abilities and personality traits) significantly influence visualization processing, in a way that warrants providing support adapted to these user differences. For instance, low levels of the cognitive abilities *perceptual speed* and *visual working memory* (WM) have been linked to lower performance in simple analytic tasks with bar-chart-based visualizations [1], [27]. Low levels of *spatial abilities* have been linked to worse performance in map reading tasks [28] and probabilistic reasoning tasks performed with icon array visualizations [29]. Low *vis literacy* was found to hinder user experience during decision making tasks supported by maps and deviation charts [4], as well as during processing network visualizations in science museums [30]. Users with low levels *of need for cognition*, a personality trait, obtained worse performance than their counterparts in low-level analytical tasks with colored boxes [31]. *Locus of control*, another personality trait, was shown to impact whether a user performs best with tree-based or box-based visualizations [2], [32].

 Several long-term user characteristics were also linked to user performance during MSNV processing. Toker et al. [23] tested the influence of nine user characteristics on user performance when completing the task of reading and answering comprehension questions about bar-chart-based MSNVs extracted from real-world sources (e.g., newspapers). Results showed that users with low levels of *reading proficiency* and *verbal WM* were significantly slower in task completion than users with high levels of these abilities. Users with low levels of *vis literacy*, *need for cognition* and *verbal IQ* were significantly less accurate on comprehension questions than their counterparts. Analysis of users' eye-tracking data showed that these worse performances were due in part to difficulties in identifying referenced data points in the MNSV visualizations. In this paper, we leverage the same set of MSNVs used in [23] to evaluate the usefulness of gaze-driven interventions that dynamically highlight referenced data points. We also examine if the five user characteristics found in [23] to impact performance influence the effectiveness of the interventions.

### 2.2 Eye-Tracking for User Adaptation

Existing research has investigated the potential of eye-tracking to support adaptation by detecting in real-time relevant user gaze patterns, *short-term* states and *long-term* user characteristics.

Gaze-driven adaptation, which reacts to specific user gaze patterns, has been investigated in several domains. For instance, in educational settings gaze-driven prompts were used to: (i) refocus student attention back to the screen when they look away while interacting with educational software, with positive results on student learning [33]; (ii) assist children during reading by pronouncing out loud words that they fixate for a long time [34]. Shirazi et al. [35] adapted online advertisements based on what information users look at in e-commerce webpages. In InfoVis, gaze-driven adaptation has been studied to support map processing [14], [15]. Göbel et al. [14] dynamically place the legend of a map next to where the user is looking, and highlight in the legend the symbols that lie in the area of gaze location, with positive results for both processing time and user satisfaction. Bektas et al. [15] deemphasize the parts of a map that are outside the user's focus of attention, but this adaptation made no difference is user performance in simple visual search tasks.

Research has shown that eye-tracking can reveal more about the users than where they look. In particular, eye-tracking data has been used to predict user short-term states such boredom [36] and mind wondering [37] in educational settings, as well as interest [11], confusion [12] and cognitive load [13] during visualization processing. Still in the context of visualization processing, there are results on leveraging eye-tracking to predict long-term cognitive abilities relevant for adaptation, such as perceptual speed, verbal WM, visual WM, visual scanning [8]–[10]. These results show that eye-tracking can reveal rich information about the users. However, thus far only the work by D'Mello et al. [37] on predicting mind wandering while studying educational text has been used to drive adaptation, in this case prompting user to refocus their attention.

### 2.3 Cuing for Multimodal Documents

Cuing, or adding visual prompts that guide user attention, has been extensively investigated to facilitate the processing of multimodal instructional material consisting of text and accompanying diagrams or pictures (but not visualizations, see [20] for an overview). In particular, color coding matching parts of the text and the graphics was found to increase comprehension [38]–[40]. This color guidance was provided either upfront [38], [39], or at the user request when clicking on a specific paragraph [40].

Cuing for supporting the processing of MSNV documents as been investigated in [19], by displaying all visual cues upfront in the document. The visual cues consisted of colored lines drawn over the document to link words in the text to the corresponding information in the visualization. The cues were evaluated in a task requiring users to seek specific information (targets) in MSNVs. The targets were predefined so that the linking could be provided upfront, and the number of targets was limited to avoid clutter. Results show that this form of cuing reduced search time. However, providing all cues upfront it is hard to scale to MSNVs with a large number of references, as it is often the case in real-world documents[1], because the many cues can visually clutter the document and create overlaps, thus diminishing the effectiveness and readability of the visualization [41], [42].

The form of gaze-driven cuing we adopted in this paper was originally proposed in [22], and is further inspired by a study that compared different ways of highlighting relevant bars in stand-alone bar charts (i.e., not embedded in MSNVs) [43]. We chose to rely on this work because it explicitly investigated cuing initiated dynamically as opposed to provided upfront (as done in [39], [40] and [19]) or upon specific user request (as done in [41]). The study tested various forms of highlighting interventions, including thickening the border of the relevant bars, deemphasizing non-relevant bars, and drawing connecting arrows or reference lines. These interventions were tested in a series of low-level analytical tasks. For each task, the user would see a bar graph and a question asking to retrieve and compare specific data points, with the bars corresponding to these data points being highlighted after a short time delay. This delay was added to mimic the effects of receiving cuing dynamically during visualization processing and ascertain if it could be annoying or distracting. Results showed that users performed significantly better when receiving the thickening and deemphasizing interventions, compared to receiving none, providing preliminary evidence on the potential of dynamic cuing for bar charts processing.

## 3 ADAPTIVE MSNV

In this section, we first describe the MSNVs that we leverage to evaluate our proposed gaze-driven highlighting interventions. Then, we describe the design of these interventions and their implementation for the target MSNVs.

### 3.1 MSNV Dataset

We used a set of 14 bar-chart-based MSNVs that Toker et al. [23] derived from an existing dataset of 40 MSNVs extracted from real-world sources, e.g., Pew Research, The Guardian, The Economist [21]. To keep the study complexity manageable, these MSNVs feature only bar charts, one of the most commonly used visualizations in real-world documents [17]. The references in this dataset had been previously identified via a rigorous coding process, indicating which data points in each visualization correspond to each sentence(s), as detailed in Kong et al. [21]. Each MSNV in this dataset consisted of "snippets" of larger source documents whereby each snippet included a self-contained excerpt of the original text and one accompanying bar chart. We use this format in order to more easily control for different factors of complexity of the MSNVs that might impact their processing. In particular the 14 MSNVs were selected to include a balanced variety of bar chart types (simple, stacked, grouped), length (measured in terms of number of words and references) and number of referenced data points. Figure 2 shows two MSNVs with different complexity, whereas Table 1 shows summary statistics on the composition of the 14 MSNVs. The selection process is fully described in [23].

---

[1] Some of the MSNVs we use in our study come from Pew Research documents on public policy that can include up to 30 references [21].

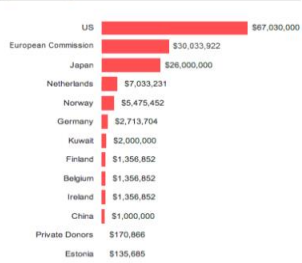
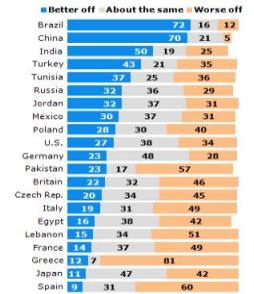

Fig. 2. Two MSNVs with different levels of complexity: *(i)* the one on the left with one reference (dashed underlines, added to this figure for clarity) and a simple bar charts, *(ii)* one on the right with three references and a stacked bar charts showing many more data points.

TABLE 1
SUMMARY STATISTICS FOR THE PROPERTIES OF THE MSNVS IN THE USER STUDY

| MSNV Property | Min | Max | Mdn | Mean | SD |
|---|---|---|---|---|---|
| **Total number of words** in narrative text | 43 | 228 | 75 | 90.8 | 49.7 |
| **Total number of sentences** in narrative text | 2 | 14 | 4 | 4.9 | 3.0 |
| **Total number of references** in narrative text | 1 | 7 | 3 | 2.8 | 1.8 |
| **Total number of data points** in visualization | 4 | 63 | 14 | 22.1 | 19.7 |
| **Total number of referenced data points** | 2 | 24 | 6 | 10.1 | 7.8 |

### 3.2 Design of the gaze-driven highlighting

Our proposed gaze-driven interventions dynamically highlight those bars in an MSNV chart corresponding to the reference the user is reading, as detected by eye-tracking[2]. The rationale for this form of guidance is to drive users' attention to the appropriate data in the charts when it is most relevant, i.e., when the user is attending to that piece of information in the text.

Designing these highlighting interventions entails several challenges related to determining the main properties of the interventions, namely: *(i)* what type of highlighting to use; *(ii)* when exactly to trigger the intervention during the reading of a reference; *(iii)* whether interventions should be incrementally added to the bar chart as references are read, or whether previous interventions should be removed so that only one is active at any given time.

Testing values for all these properties in a formal study is not feasible as it would generate too many study conditions. Instead, we conducted dedicated pilot studies where we collected feedback on a minimal set of suitable values for these three properties and either identified a clear winner to use moving forward, or alternatives to test in the formal study. In these pilots, users read the same set of 14 MSNVs and completed the same task as in the main study (described in Section 4.1), with various versions of the adaptive interventions based on the property values we wanted to test. Participants were then interviewed to elicit their preferences and feedback.

**Highlighting type.** For this property, we chose to pilot test the two types of highlighting found to be most effective at supporting bar chart processing in [43] (see related work): thickening the border of the relevant bars, and desaturating non-relevant bars (see Figure 3 for examples).

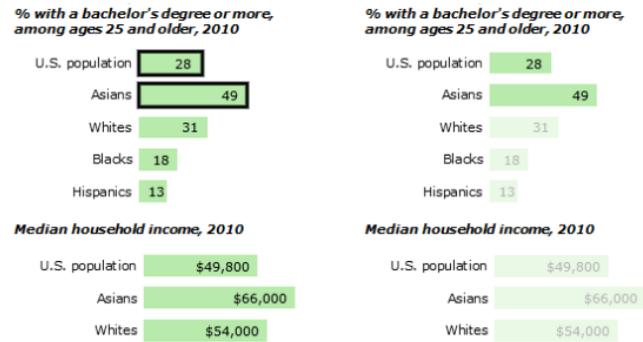

Fig. 3: Sample highlighting of the top two bars, via thickening of their borders (left), and desaturating the other bars (right).

Five out of 6 pilot users reported that desaturating bars was too disruptive, because it removes context by making it difficult to see the desaturated bars even if they wanted to. Based on this feedback we retained *thickening* of the border of the bars for the study. The borders are always thickened in black so as to use a neutral color that has no other visual encoding in the bar charts of the MSNVs, as done in [43]. To ensure that the black outlines are noticeable for all the bars in the dataset, we adjusted the brightness and saturation of all bar colors so that their contrast ratio with the black outline is consistent across MSNVs. We also ensured that the colors remain well visible and distinguishable.

**Intervention timing.** Because we want to trigger interventions when a user is reading references in the text, we use an eye tracker to track the user's fixations[3] over these references. An intervention for a reference is then triggered whenever a *sufficient* number of fixations on the related sentences have been detected. An open question, however, is when exactly to trigger the intervention during the reading of a reference, e.g., at the start of the sentence, when the sentence has been fully read, somewhere in between.

For this study, we chose to test the option of triggering

---

[2] A video of the interventions is available at *http://github.com/ATUAV/ATUAV_Experimenter_Platform/blob/master/documentation/ATUAV Experimenter platform - Demo video with sample adaptation.mp4*

[3] A fixations is defined as gaze maintained at one point of the screen for at least 100ms, as captured by the Tobii eye-tracker.

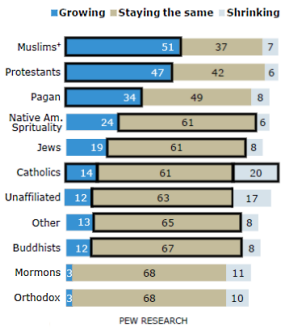 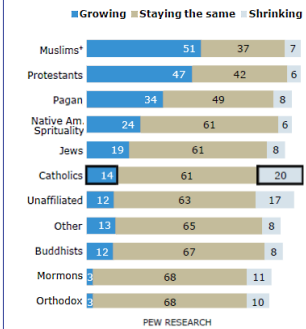 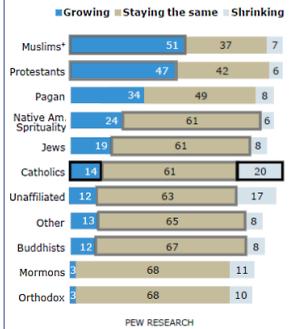

Fi. 4. Example removal strategies. The 4 references underlined in the text have been read, with the one at the bottom being the current one active. (A) keep all highlight; (B) remove previous highlighting; (C) desaturate previous highlighting in grey.

the interventions when the user has read more than half of the sentence, so that they have sufficient context to process the relevant bars in the chart[4]. The challenge here is that the number of fixations needed to process a reference can depend on its length, difficulty or phrasing, as well as on the reading speed of each user. We leveraged the eye-tracking data collected in [23], which used the same dataset as we use here, to compute the average number of fixations users spent on each reference in our target MSNVs. This average ranges from 8 for the shortest reference to 45 for the longest one (overall mean = 24, st. dev. = 10).

Next, we needed to define which percentage of this average number of fixations per reference should be considered as sufficient for having processed the reference and for triggering the intervention. Choosing a high percentage is risky because it is prone to triggering interventions too late, when the user has finished reading the reference and has moved to subsequent text. This is especially true for fast readers or readers who skim through the text, as they would generate fewer than average fixations on a sentence.

We chose to first pilot a trigger threshold of 60%, i.e., for each reference, the corresponding intervention is triggered when 60% of the average fixations required to read it is detected. Five out of 6 pilot users reported that the interventions seemed to appear too late (e.g., "*It feels like there is a delay*"). We also noticed that 3 of these users each triggered less than 70% of the interventions, either because they were reading fast and thus did not generate enough fixations over the reference, or because our 60% threshold was generally too conservative. Based on this feedback we lowered the trigger threshold to 40%, to better ensure that the interventions would be triggered, even by faster readers. Although for some users this threshold might trigger interventions when they are just partway through reading a sentence, for our purposes it is important to make sure that as many interventions as possible are delivered, given the objective to investigate the effectiveness of these interventions compared to not receiving them. We tested this 40% threshold with two additional pilot users, who triggered most interventions and found their timing to be suitable, thus we retained this threshold for the study.

**Intervention removal.** Since most MSNVs contain multiple references, we had to determine whether previously triggered interventions should be left active or removed upon the delivery of a new one. Leaving all interventions active is shown in Fig. 4 (A), where four references (underlined) were read, of which the current one is at the bottom, and all the corresponding bars are all highlighted. This approach facilitates going back to the previous references, which can be useful if the user forgot some information, or want to compare data points across references. A possible drawback is that having too many highlighted bars might become overwhelming, and might make it difficult to discern the bars related to the current reference read (e.g., the 'Catholics' bars in Fig. 4 (A), corresponding to the reference at the end of the text), especially in documents with many references.

An alternative strategy is to remove all previous interventions, as shown in Fig. 4 (B), where only the 'Catholics' bars (i.e., the latest intervention) are left highlighted. With this approach the user can easily focus on the most current intervention, but cannot refer to previous ones anymore.

Pilot testing both strategies with 6 users revealed mixed feelings, with no clear winner. Four users reported that removing previous highlighting was unhelpful because they could not remember what bars they already processed. Two of them also said they often had to go back and re-read the text due to that. Three pilot users liked having all the interventions after reading the entire text because it provides "*a good visual summary of the salient information described in the MSNV*". However, five out of 6 users reported difficulties in distinguishing between the previously highlighted bars and the recent ones with this strategy.

To account for this feedback we implemented a third strategy designed to leverage the pros of the previous two without the drawbacks. This strategy involves keeping previous highlights but desaturating the thickening so that the black outline become grey (see Figure 4, C), thus distinguishing the most recent highlighting from the previous ones. We pilot tested this strategy with two additional users, who provided very positive feedback. Thus, we retained it for the main study.

---

[4] The opposite approach is to trigger the intervention as soon as the user starts reading the sentence, but we deemed this to be potentially too distracting, as well as prone to error because the interventions might be triggered by a few spurious or slightly inaccurate fixations.

### 3.3 Platform to test Adaptive MSNV

We have developed a dedicated platform to generate, deliver and evaluate interventions in user-adaptive visualizations, including the gaze-driven interventions for MNSV described in this paper. A full description of the platform is provided in its user manual[5]. Here we provide a high-level description for the purpose of this paper.

The platform is built upon Tornado (*www.tornadoweb.org*), a Python web framework for developing asynchronous client-server applications. The server side of the platform is responsible for both processing eye-tracking data in real time (*back end* component), as well as triggering adaptation (*middle end* component). Specifically, the *back end* establishes connection with the eye-tracker, fetches the raw data at the eye-tracker's frequency, and processes it to extract higher-level information that can be used to trigger adaptation. This includes fixations, saccades (eye movements between two fixations), pupil sizes, distance of the eyes to the screen, and a set of eye-tracking features that capture a user's attentional patterns both over the entire screen as well as within pre-specified Areas of Interests (AOIs). For this study, the AOIs of interest are the individual reference sentences in each MNSV in our dataset, and the platform tracks user fixations over these AOIs[6].

The *middle end* manages the delivery of adaptation by evaluating a set of *adaptation rules* over the livestream of eye-tracking features generated by the back end. These rules are implemented in SQL, and experimenters can quickly add new rules in the platform to test new forms of adaptive interventions in their visualizations. For this MSNV study, there is one general rule that initiates an intervention when the number of detected fixations over a given reference exceeds its triggering threshold (see Section 3.2). The middle end also provides the necessary functionalities to run a user study and evaluate the adaptation, including data logging, task randomization, and the display of questionnaires and tests as needed.

The client side of the platform is responsible for displaying the target visualization (in our case the MSNVs) and the adaptive interventions upon notification from the middle end. Interventions are displayed in the visualizations via JavaScript callbacks, and we use the D3 and Angular JavaScript libraries to plot the highlighting.

## 4 USER STUDY

To evaluate the gaze-driven adaptive highlighting for MNSVs described in the previous section, we used a *between-subject design*, where we compare the performance of a group of users who read MNSV with the highlighting interventions (*adaptive group*) and a *control group* that reads the same MSNV with no highlighting. The data for the control group come from the study reported in [23], referred to as *control study* from now on. The data for the adaptive group comes from the study we describe in the rest of this section (*adaptive study* from now on). The two studies use the exact same task and procedure, fully described in [23] and summarized in Section 4.1.

### 4.1 Participants and Study Procedure

A total of 119 subjects were recruited in the studies via advertising at our campus and on Craigslist, and were paid $35 for participating. The control study [23] included 56 subjects (32 female), with age from 19 to 69 (M=28, SD=11). For the adaptive study we recruited 63 participants (34 female), with age from 18 to 59 (M=25, SD=8). In both studies, about 60% of participants were university students, and the others were from a variety of backgrounds (e.g., retail manager, restaurant server, artist, nurse, retired).

The study procedure involves a single session lasting at most 90 minutes. The session starts with the participant undergoing calibration with the eye-tracker, a Tobii T-120. high-end camera-based remote eye-tracker embedded in a display of 1280 x 1024 pixels, with sampling rate of 120 Hertz. Next, participants are given the task of reading a MSNV document on the computer screen, and signal when they are done by clicking 'next'. At this point, they see a screen with a set of questions that elicit their opinion of the document and test their comprehension (see section 4.3). Participants perform this same task for the 14 MSNVs described in Section 3.1. The ordering of the 14 MSNVs is randomized for each participant. Participants are not given a time-limit to read the MSNVs, nor training on the interventions, to mimic how they might be used in realistic settings. To ensure that participants dedicated adequate effort to the task, a $50 bonus was promised for the three participants with the best performance, evaluated in terms of both speed and accuracy. In the adaptive study, participants also filled out a postquestionnaire to rate the usefulness and ease of use of the adaptation (Section 4.3). They were also briefly interviewed to discuss their ratings.

Standard tests from psychology were used to assess, for each participant, a battery of user characteristics (described in the next section) that might influence how MNSV are processed and how participants react to adaptive interventions. To reduce fatigue, some of these tests, which are computer-based and do not require an invigilator, were given to participants to do at home prior to the experiment. The rest, which were either paper-based or required specialized software not available remotely, were administered during the study session. See [23] for more details on tests administration procedure, which was kept identical in the adaptive study.

### 4.2 User Characteristics

Nine user characteristics were measured in both the control and the adaptive study, to keep the same study procedure. As we discussed in the related work, [23] only five out of these nine were found to influence user performance when processing MSNVs, i.e., low levels of these characteristics were linked to worse performance. Because of their link to performance, we focus on these five for our analysis of how user characteristics might influence the effectiveness of adaptive interventions for MSNV.

---

[5] github.com/ATUAV/ATUAV_Experimenter_Platform/tree/master/documentation. All code is available at https://github.com/ATUAV/ATUAV_Experimenter_Platform/

[6] We used a JS script (included in the platform on Github) to automatically extract the AOIs coordinates from the documents.

TABLE 2
SET OF TESTED USER CHARACTERISTICS, AND SUMMARY STATISTICS OF THE PARTICIPANTS' SCORES IN BOTH STUDIES

| User Char. | Definition | Instrument | Score range |
|---|---|---|---|
| VIS LITERACY | Ability to use a visualization to translate questions specified in the data domain into visual queries in the visual domain, as well as interpreting visual patterns in the visual domain as properties in the data domain [24]. | Visualization Literacy 101 – Bar Chart Test [24] | -2 ; 2 |
| VERBAL WM | Quantity of verbal information (e.g., words) that can be temporarily maintained and manipulated in working memory [43]. | OSPAN (Operation-word span) [44] | 0 ; 6 |
| READING PROFICIENCY | Vocabulary size and reading comprehension ability in English [45]. | X_Lex Vocabulary Test [45] | 0 ; 100 |
| VERBAL IQ | Overall verbal intellectual abilities that measures acquired knowledge, verbal reasoning, and attention to verbal materials [46]. | North American Adult Reading Test [46] | 75 ; 125 |
| NEED FOR COGNITION | Extent to which individuals are inclined towards effortful cognitive activities [47]. | Need for Cognition Scale [47] | -36 ; 36 |

These five user characteristics are defined in Table 2, along with the standard tests from psychology that were used to measure them. The first characteristic (vis literacy) relates to how well users can process visualization; the next three (verbal WM, reading proficiency, and verbal IQ) relate to the ability to process textual elements; finally, need for cognition is a personality trait defining how much users like effortful cognitive activities[7].

### 4.3 Dependent Measures

Two sets of dependent measures used in the adaptive study were the same as those collected in [23] and relate to *task performance* and *user experience*. In addition, in the adaptive study we collected measures related to the user *perception* of the interventions.

Fig. 5. Questions presented to users after reading each MSNV.

**(i) Task performance.** This set comprises of *time on task* and *accuracy* in processing each MSNV. Time on task is the time elapsed between when a participant is shown an MSNV, and when they signal that they are done reading it by hitting "next". Accuracy is the ratio of correct answers to the comprehension questions that appear after pressing "next". These were adapted from [44] and include two *recognition questions* asking to recall specific information from the MSNV (questions 3-4 in Fig. 5), and a *title question* asking to select a suitable alternative title for the MSNV (question 5 in Fig. 5) as a way to test overall comprehension. See [23] for more details on the question design.

**(ii) User experience.** This set comprise two subjective measures about the perceived *ease-of-understanding* and *interest* of the document, assessed on a 5-point Likert scale (questions 1-2 in Fig. 5).

**(iii) Perception of the interventions.** This set comprises 10 measures related to user perceived *usefulness*, *ease-of-use* and *satisfaction* with the interventions. These measures were collected via a web questionnaire displayed after a participant read all 14 MSNVs, where the participant rated the 10 statements listed in Table 3 on a 7-point Likert scale ranging from "Strongly disagree" to "Strongly agree". Statements 1-5 ask about the perceived usefulness of the interventions, including possible negative aspects, namely if they were distracting or confusing. Statements 6-9 ask about aspects related to the ease-of-use of the interventions. The last statements gauge satisfaction with the interventions. This questionnaire was derived from the USE questionnaire [45], a well-established instrument to evaluate the usability of user interface features.

TABLE 3
QUESTIONNAIRE ON PERCEPTION OF THE INTERVENTIONS

| Component | Statements |
|---|---|
| USEFULNESS | 1. I found the interventions useful.<br>2. I found the interventions useful to understand the document.<br>3. I found the interventions useful to focus on the relevant information.<br>4. I found the interventions distracted.<br>5. I found the interventions confusing. |
| EASE OF USE | 6. I found that the timing of the intervention was right.<br>7. I found the interventions easy to notice.<br>8. I found the intervention well-integrated into the document. |
| SATISFACTION | 9. I was satisfied with the intervention.<br>10. I would use the interventions in my daily life. |

---

[7] As done in [23], we verified that these five measures are not correlated, as per a Kendall rank correlation test (τ < 0.30).

## 5 ANALYSIS OF THE TRIGGERING MECHANISM

To make sure that the intervention triggering mechanism discussed in section 3.2 generated enough interventions to answer our research question, we examined the percentage of interventions each participant triggered, given the maximum number of 35 available. Fig. 6 reports these percentages. It should be noted that we discarded 5 users because of too many invalid gaze samples (as reported by the eye-tracker), leaving 58 users for this and subsequent analyses.

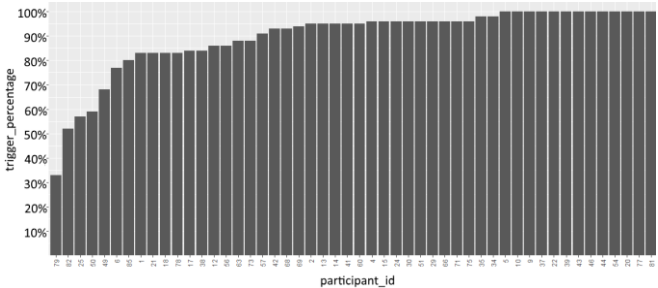

Fig. 6. Percentage of triggered interventions per participants.

On average, these 58 participants triggered 81% of the interventions (SD=19%), and Fig. 6 shows that a large majority triggered most of the interventions. Specifically, 46 users (about 79%) triggered more than 75% of the interventions. The fact that not all interventions were triggered is to be expected, as it is a normal reading behavior to skim or even skip some sentences at times, for example when a text is not of interest to the participant.

For the 13 participants who triggered less than 75% of the interventions, we investigated whether this was due to problems with eye-tracking, or to them being fast/skim readers. To do so, we looked at the gaze heat map for each of these users, which can reveal calibration issues if the heat map is consistently misaligned with the sentences in the text and the bar chart.

The heat map of 8 of these 13 participants showed such misalignment (see example on Fig. 7), preventing the triggering of interventions when the eye-tracker would not detect fixations over reference sentences that the participant actually read, or alternatively triggering an intervention for references that the user has not read yet. Because this technical issue strongly affected the delivery of the interventions, we discarded these users from further analysis.

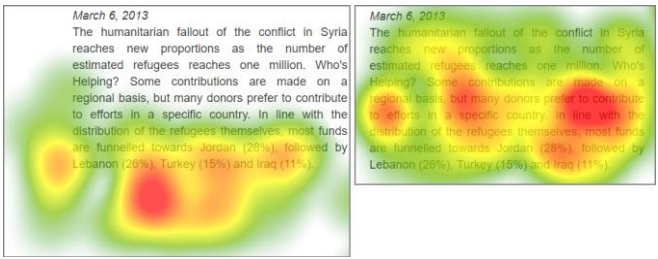

Fig. 7. Heat map misaligned with the text (left) vs an aligned one (right)

There was no obvious issue with the heat maps of the 5 remaining participants who triggered less than 75% of interventions, and their low trigger percentage is likely due to a high reading speed or a tendency to skim the text. Thus, we retained these users, remaining with 50 participants who, on average, triggered 89% of the interventions (st. dev. = 14%). We consider the proportion of interventions triggered by these 50 users to be adequate to proceed with our analysis. However because there is some variance in their trigger percentage, we will discuss if/how it affects outcomes for the adaptive group in the results section.

## 6 ANALYSIS AND RESULTS

The first part of this section (Section 6.1), presents the analysis and results related to our research question:

"*Does gaze-driven highlighting of relevant parts of an MSNV graph improve user performance and subjective experience with bar-chart-based MSNVs, compared to users who received no such highlighting? Do the results depend on one or more user characteristics that were previously found to influence MSNV processing?*"

The rest of the section presents further results on the effects of interventions in the adaptive group. Section 6.2 discusses how participants in the adaptive group perceived the interventions. Section 6.3 reports on how the number of interventions received influenced outcomes.

### 6.1 Comparison of Control and Adaptive Groups

#### 6.1.2 Analysis

To answer our research question we compare performance and subjective experience of the participants in the control and adaptive groups, while accounting for the possible influence of the 5 user characteristics presented in section 4.2 (UC from now on). Recall that we measure performance in terms of *time on task* and *accuracy*, and user experience in terms of perceived *ease-of-understanding* and *interest* of the MSNV (see section 4.3). Statistics for these dependent measures are shown in Table 4.

TABLE 4
SUMMARY STATISTICS OF DEPENDENT MEASURES.

| Measure | Control | Adaptive |
|---|---|---|
| ACCURACY (%) | 71.9 (±30) | 74.4 (±31) |
| TIME-ON-TASK (SECS) | 56.3 (±32) | 60.1 (±33) |
| INTEREST | 3.37 (±1) | 3.31 (±1) |
| EASE-OF-UNDERSTANDING | 4.00 (±1.2) | 4.05 (±1.2) |

We ran four mixed models, one per dependent measure with group (adaptive, control) as independent variables, and the five UC as covariates. Participant ID and MSNV ID were added as random effects in the mixed models, to account for variability across the participants and the documents, respectively. Each mixed model was fitted with a bidirectional stepwise algorithm for model selection based on AIC, using the lmerTest package in R [46]. To account for family-wise error, resulting *p*-values from all four mixed models are adjusted using the Benjamin–Hochberg procedure to control for the false discovery rate (FDR) [47].

#### 6.1.2 Results

Since we are interested in the impact of having or not having adaptive interventions, we focus on results pertaining to effects involving groups. We found no significant[8] main effect of group on the dependent variables. However, there is a significant interaction effect of *vis literacy* with *group* on *accuracy* ($F_{1,97}$ = 12.71, $p$ = .0006, $r$ = 0.34).

To analyze the directionality of this interaction effect, we divide all participants into three bins based on a 3-way split of their vis literacy scores (Low, Medium, High). Figure 8 shows the interaction effect between *vis literacy* with *group* on *accuracy* using this three way split.

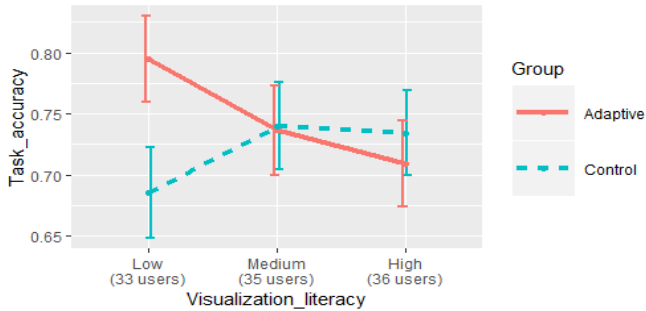

Fig. 8. Interaction effect of vis literacy with group on accuracy. Error bars show 95% confidence intervals.

Post-hoc FDR adjusted pairwise comparisons show two significant effects in this interaction:

*(i)* Users with *low levels of vis literacy* were *more accurate* in the adaptive group than in the control group ($p$ = .0009, $r$ = .33), with a substantial boost in accuracy of nearly 11% on average (see Figure 8, left). It should be noted that the mixed models revealed no significant interaction effect of vis literacy and group on time on task ($p$ = .19, $r$ = .15), indicating that the increased accuracy of the low vis literacy users with the adaptive interventions did not come at the expense of longer time on task.

*(ii)* In the adaptive condition, *low vis literacy* users were significantly ($p$ = .006, $r$ = .28) *more accurate* that *high vis literacy* users (see plain orange line in Figure 8), with an improvement in accuracy of about 9%.

### 6.1.3 Discussion

The positive effect of interventions for low vis literacy users is noteworthy because previous studies (e.g., [4], [23], [30], [48]) have confirmed that having low vis literacy creates a disadvantage when working with visualization. Our finding provides promising initial evidence that gaze-driven highlighting interventions can help alleviate such disadvantage in the context of MSNV processing, so much so that low vis literary users outperformed in accuracy the high vis literacy ones in the presence of the interventions, without taking longer time on task.

High vis literacy users in our study were neither helped nor harmed by the interventions, as indicated by the lack of significant difference in accuracy ($p$ = 0.567, $r$ = 0.08) and time on task ($p$ = 0.954, $r$ = 0.04) for these users in the control and adaptive group. However, these users did not achieve a ceiling effect in accuracy (in fact they were outperformed by low literacy users in the adaptive group), indicating that it is worthwhile to explore other forms of guidance to help them process MSNVs more effectively.

## 6.2 Perception of the adaptive interventions

### 6.2.1 Analysis and Results

We analyze perception of interventions for participants in the adaptive group via the ratings that these users provided for the 10 statements related to usefulness, ease-of-use and satisfaction, described in section 4.3. Because the ratings for some of these statements were highly correlated, we selected only the four most distinct measures, namely those related to perceiving the interventions as *useful*, *delivered timely*, *confusing* and *distracting*. Table 5 provides summary statistics for these measures.

Overall, the participants' ratings were positive for *useful*, *timing* and *confusion*, with modes of respectively 5 ("somewhat useful"), 6 ("delivered at the right time",) and 2.5 (between "not confusing" and "somewhat not confusing"). However, participants found the interventions to be "somewhat distracting" (mode of 5).

TABLE 5
SUMMARY STATISTICS OF SUBJECTIVE PERCEPTION OF THE INTERVENTIONS ON LIKERT SCALES RANGING FROM 1 TO 7

| Measure | Mean | Mode | Sd | Min | Max |
|---|---|---|---|---|---|
| USEFUL | 4.51 | 5 | 1.54 | 1 | 7 |
| TIMELY DELIVERY | 4.78 | 6 | 1.57 | 1 | 7 |
| CONFUSING | 3.07 | 2.5 | 1.66 | 1 | 7 |
| DISTRACTING | 4.25 | 5 | 1.83 | 1 | 7 |

### 6.2.2 Discussion

The high rating for intervention timing indicates that our chosen threshold for triggering interventions (cf. section 3.2) was suitable. The rather low rating for confusion is especially noteworthy because participants received no training with the interventions prior to the study tasks, thus they could have misunderstood their meaning or the reasons for their appearance. The low levels of perceived confusion suggest that this was not a major factor, and that the interventions were intuitive enough for most participants. It is possible, however, that introducing the users to the adaptive interventions beforehand might further reduce confusion ratings. The good ratings for confusion and timing also suggest that there were not many instances of interventions wrongly triggered because the eye-tracker detected spurious fixations on the corresponding references when the user was not reading them. However, a detailed analysis of the eye-tracking logs is needed to ascertain if and when these events happened.

The fact that participants found the interventions to be somewhat distracting is to be expected given that the interventions are provided dynamically during reading. Nonetheless, the levels of perceived distraction remain moderate on average, and participants still reported that they found the interventions to be useful despite this distraction. Furthermore, the distraction did not escalate into confusion. Still, moving forward it will be important to investigate the specific reasons for distraction and ways to mitigate them.

---

[8] We report statistical significance at the .05 level, as well as effect sizes as high for $r$ > .5, medium for $r$ > .3, and low otherwise.

## 6.3 Impact of Percentage of interventions Received

### 6.3.1 Analysis and results

As discussed in section 5, there were differences in the percentage of interventions triggered by each participant in the adaptive group (mean = 89%, st. dev. = 14%, min = 31%, max = 100%). Here, we investigate whether these differences influence user performance, experience, and perception of the interventions. To facilitate the analysis, we discretized the number of received interventions into two bins (High and Low) via a median split[9]. Users in the Low bin received 77% of the interventions on average (st. dev. = 17%, min = 31%, max = 88%), and users in the High bin 97% on average (st. dev. = 2%, min = 91%, max = 100%).

For each of the four measures of performance and experience (Table 4), we run a mixed model with the percentage of interventions triggered (Low or High) as the factor, and participant ID and MSNV ID as random effects. For each of the four measures of intervention perception in Table 5, we run a Kruskal-Wallis test with percentage of interventions triggered as the independent variable (we use this test because perception was rated once, at the end of the study, therefore it is not a repeated measure as performance and experience are). All results are adjusted using the Benjamin–Hochberg procedure.

We found significant main effects of percentage of interventions triggered on: *(i)* time-on-task ($F_{1,171}$ = 74.97, $p < .0001$, $r = .55$); *(ii)* distraction ($x^2_{(1)}$ = 6.77, $p < .009$, $r = .35$); confusion ($x^2_{(1)}$ = 5.53, $p < .019$, $r = .38$). The directionality of these effects indicates that participants who triggered fewer interventions had shorter task times, and reported more confusion and more distraction.

### 6.3.2 Discussion

The fact that users who triggered fewer interventions had shorter times on task might indicate that receiving interventions slow users down. However, as discussed in Section 6.1.2, we found no difference in time on task between the adaptive group and the control group ($p = .19$, $r = .15$), which contradicts this hypothesis. Another possible explanation is that users who have a shorter time on task were fast readers who, because of their faster reading speed, happened to trigger fewer interventions.

The finding that users who received fewer interventions were significantly more distracted might be due to the fact that interventions were triggered more erratically. Their higher confusion might be due to the fact users expected some important information to be highlighted in the bar charts, whereas it was not always the case since they received less highlighting overall. The finding that users who received fewer interventions were significantly more distracted might be due to the fact that interventions were triggered more erratically. Their higher confusion might be due to the fact users expected some important information to be highlighted in the bar charts, whereas it was not always the case since they received less highlighting overall.

## 7 CONCLUSION

We conducted a user study aimed at evaluating the value of gaze-driven highlighting interventions in Magazine-Style Narrative Visualizations (MSNVs), i.e., visualizations embedded in narrative text. Specifically, we leverage eye-tracking to detect in real time when the user is reading a sentence describing specified data points in the visualization, so as to dynamically highlight these data points in the visualization. We compared the performance and subjective experience of participants who received the gaze-driven highlighting against a control group who received none, and we also examined if users' performance and experience were influenced by long-term user characteristics known to play a significant role during MSNV processing. Our results show that the interventions were overall well perceived by the users and they benefitted specifically to users with low levels of vis literacy, who have lower ability in processing and understanding data visualizations.

In this study we made two key contributions. First, we broadened the evidence on the value of eye-tracking for adaptive visualizations. Specifically, our study is the first to show the value of gaze-driven adaptation with bar charts embedded into MSNVs, whereas previous work only studied gaze-driven adaptation in map-based visualizations with no narrative text. This finding is important because bar charts are popular, commonly used visualizations, and MSNVs constitute a widespread context of usage of information visualizations in real-world media (press, internet, scientific publications).

A second contribution is that our results are the first to show that long-term user characteristics can influence the effectiveness of gaze-driven adaptation in visualization. In particular, the fact that the interventions helped users with low levels of *vis literacy* is significant, because it shows that the interventions benefited those users who needed help the most due to their lower abilities. Our results also show that users with high levels of vis literacy were not impacted by the interventions, but were overall outperformed by low literacy users who received them. Thus, despite their higher vis literacy, these users did not show ceiling performance and might benefit from other forms of adaptive interventions more suitable to their specific needs. We plan to start investigating this point by analyzing the feedback provided by these high vis literacy users during the post-interviews we conducted at the end of each session, as this feedback might provide specific ideas on how to make the interventions more suitable to them.

This research is a first step towards designing personalized gaze-driven support for MSNV processing. As such, it provides proof of concept for the potential of this guidance, but also has several limitations to be addressed in future work, as we discuss next.

The documents used in the study were excerpts from real-world MSNVs, usually quite shorter than the original documents, and we do not know how our results would generalize to these. We argue that the type of gaze-driven guidance we investigate should be even more helpful in longer, more challenging documents. However, it is possible that some of our findings on which intervention properties worked well (e.g., for type of highlighting, removal strategies etc.) might have to be adjusted. We plan to address this point by running a new study focusing on testing the proposed interventions with full length MSNV.

---

[9] Keeping the continuous values or discretize into more than two bins is difficult because of the sparseness for low values.

Although users overall appreciated the interventions, i.e., found them useful, delivered with good timing, and not that confusing, these scores still had room for improvement. Also, users found the interventions to be somewhat distracting (although this distraction did not appear to hinder performance or user experience). We plan to examine how to reduce distraction and further improve the other scores of user perception, starting with the analysis of the study post-interviews to identify specific feedback that we can use to improve the interventions design and delivery.

Although participants were able to trigger most interventions (about 90% of them on average), our gaze-driven interventions are quite sensitive to a user's reading speed, but currently are not calibrated to it, causing some users to not trigger all the available interventions, which in turns can increase distraction and confusion. We plan to investigate how to include this calibration in a way that is realistic for real-world settings, so that the timing of the interventions can be personalized to each user's reading speed.

There are several other steps of future work on our agenda, which will be facilitated by the availability of the software platform we have devised to support the implementation and the evaluation of eye-tracking-based adaptive support for visualization processing. For instance, we will experiment with adding cuing that guides the user attention from the visualization back to the relevant reference in the text (i.e., the relevant references). We will compare our highlighting interventions with other types of dynamic cuing, i.e., a gaze-driven version of the static linking interventions presented in [19]. We also plan to investigate gaze-driven adaptation in MSNVs with different visualizations than bar charts. Altogether, this platform expands research capabilities in studying user-adaptive visualizations, with the long-term goal of better understanding, what forms of adaptation can improve the users' performance and experience across visualizations and tasks.

## REFERENCES

[1] D. Toker, C. Conati, G. Carenini, and Mona Haraty, "Towards Adaptive Information Visualization: On the Influence of User Characteristics," *Proc. 20th Int. Conf. User Modeling, Adaptation, and Personalization*, Montréal, Canada, pp. 274–285, 2012.

[2] A. Ottley, H. Yang, and R. Chang, "Personality as a Predictor of User Strategy: How Locus of Control Affects Search Strategies on Tree Visualizations," *Proc. 33rd ACM Ann. Conf. Human Factors in Computing Systems*, Seoul, Korea, pp. 3251–3254, 2015.

[3] C. Ziemkiewicz, A. Ottley, R. J. Crouser, K. Chauncey, S. L. Su, and R. Chang, "Understanding Visualization by Understanding Individual Users," *IEEE Comput. Graph. Appl.*, vol. 32, no. 6, pp. 88–94, Nov. 2012.

[4] S. Lallé, C. Conati, and G. Carenini, "Impact of Individual Differences on User Experience with a Visualization Interface for Public Engagement," *Proc. 2nd Int. Wksh. Human Aspects in Adaptive and Personalized Interactive Environments*, Bratislava, Slovakia, pp. 247–252, 2017.

[5] D. Gotz and Z. Wen, "Behavior-driven visualization recommendation," *Proc. 14th Int. Conf. Intelligent User Interfaces*, New York, NY, USA, pp. 315–324, 2009.

[6] K. Nazemi, R. Retz, J. Bernard, J. Kohlhammer, and D. Fellner, "Adaptive Semantic Visualization for Bibliographic Entries," *Proc. Symp. Visual Computing*, Las Vegas, USA, pp. 13–24, 2013.

[7] M. Mouine and G. Lapalme, "Using Clustering to Personalize Visualization," *Proc. 16th Int. Conf. Information Visualization*, Montpellier, France, pp. 258–263, 2012.

[8] B. Steichen, G. Carenini, and C. Conati, "User-adaptive information visualization: using eye gaze data to infer visualization tasks and user cognitive abilities," *Proc. Int. Conf. Intelligent User Interfaces*, New York, NY, USA, pp. 317–328, 2013.

[9] M. J. Gingerich and C. Conati, "Constructing Models of User and Task Characteristics from Eye Gaze Data for User-Adaptive Information Highlighting," *Proc. 29th AAAI Conf. Artificial Intelligence*, Austin, TX, USA, pp. 1728–1734, 2015.

[10] C. Conati, S. Lallé, M. A. Rahman, and D. Toker, "Further Results on Predicting Cognitive Abilities for Adaptive Visualizations," *Proc. 26th Int. Jt. Conf. Artificial Intelligence*, Melbourne, Australia, pp. 1568–1574, 2017.

[11] Y.-M. Jang, R. Mallipeddi, and M. Lee, "Identification of human implicit visual search intention based on eye movement and pupillary analysis," *User Model. User-Adapt. Interact.*, vol. 24, no. 4, pp. 315–344, Oct. 2014.

[12] S. Lallé, C. Conati, and G. Carenini, "Predicting confusion in information visualization from eye tracking and interaction data," *Proc. on the 25th Int. Jt. Conf. Artificial Intelligence*, New York, NY, USA, pp. 2529–2535, 2016.

[13] S. Iqbal, P.D. Adamczyk, X.S. Zheng, and B.P. Bailey, "Towards an index of opportunity: understanding changes in mental workload during task execution," *Proc. ACM SIGCHI Conf. Human Factors in Computing Systems*, Portland, USA, pp. 311–320, 2005.

[14] F. Göbel, P. Kiefer, I. Giannopoulos, A. T. Duchowski, and M. Raubal, "Improving Map Reading with Gaze-adaptive Legends," *Proc. 2018 ACM Symp. Eye Tracking Research & Applications*, New York, NY, USA, p. 29:1–29:9, 2018.

[15] K. Bektas, A. Cöltekin, J. Krüger, and A. Duchowski, "A testbed combining visual perception models for geographic gaze contingent displays," *Proc. 17th Eurograph Conf. Visualization*, Cagliari, Italy, pp. 67–71, 2015.

[16] E. Segel and J. Heer, "Narrative Visualization: Telling Stories with Data," *IEEE Trans. Vis. Comput. Graph.*, vol. 16, no. 6, pp. 1139–1148, Nov. 2010.

[17] T. Munzner, *Visualization analysis and design*. Boca Raton: CRC Press, Taylor & Francis Group, 2015.

[18] P. Ayres and G. Cierniak, "Split-Attention Effect," in *Encyclopedia of the Sciences of Learning*, Springer, pp. 3172–3175, 2012.

[19] M. Steinberger, M. Waldner, M. Streit, A. Lex, and D. Schmalstieg, "Context-preserving visual links," *IEEE Trans. Vis. Comput. Graph.*, vol. 17, no. 12, pp. 2249–2258, 2011.

[20] T. van Gog, "The Signaling (or Cueing) Principle in Multimedia Learning," *The Cambridge Handbook of Multimedia Learning*, 2nd ed., R. Mayer, Cambridge University Press, pp. 263–278, 2014.

[21] N. Kong, M. Hearst, and M. Agrawala, "Extracting references between text and charts via crowdsourcing," *Proc. ACM Conf. Human Factors in Computing Systems*, Toronto, Canada, pp. 31–40, 2014.

[22] G. Carenini, C. Conati, E. Hoque, and B. Steichen, "User Task Adaptation in Multimedia Presentations," *Proc. 1st Int. Wksh. User-Adaptive Information Visualization*, Rome, Italy, #2, 2013.

[23] D. Toker, C. Conati, and G. Carenini, "Gaze Analysis of User Characteristics in Magazine Style Narrative Visualizations,"

*User Model. User-Adapt. Interact.*, to appear, 2019.

[24] J. Boy, R.A. Rensink, E. Bertini, and J.-D. Fekete, "A Principled Way of Assessing Visualization Literacy," *IEEE Trans. Vis. Comput. Graph.*, vol. 20, no. 12, pp. 1963–1972, 2014.

[25] E.W. Anderson, K.C. Potter, L.E. Matzen, J.F. Shepherd, G.A. Preston, and C.T. Silva, "A user study of visualization effectiveness using EEG and cognitive load," *Comput. Graph. Forum*, vol. 30, pp. 791–800, 2011.

[26] S. Lee, S.H. Kim, Y.H. Hung, H. Lam, Y. Kang, and J.S. Yi, "How do people make sense of unfamiliar visualizations?: A grounded model of novice's information visualization sensemaking," *IEEE Trans. Vis. Comput. Graph.*, vol. 22, no. 1, pp. 499–508, 2016.

[27] C. Conati, G. Carenini, E. Hoque, B. Steichen, and D. Toker, "Evaluating the impact of user characteristics and different layouts on an interactive visualization for decision making," *Comp. Graph. Forum*, vol. 33, pp. 371–380, 2014.

[28] B. Allen, "Individual differences and the conundrums of user-centered design: Two experiments," *J. Am. Soc. Inf. Sci.*, vol. 51, no. 6, pp. 508–520, 2000.

[29] A. Ottley *et al.*, "Improving Bayesian reasoning: the effects of phrasing, visualization, and spatial ability," *IEEE Trans. Vis. Comput. Graph.*, vol. 22, no. 1, pp. 529–538, 2016.

[30] K. Börner, A. Maltese, R. N. Balliet, and J. Heimlich, "Investigating aspects of data visualization literacy using 20 information visualizations and 273 science museum visitors," *Inf. Vis.*, vol. 15, no. 3, pp. 198–213, Jul. 2016.

[31] C. Conati and H. Maclaren, "Exploring the role of individual differences in information visualization," *Proc. Working Conf. Advanced Visual Interfaces*, New York, NY, USA, pp. 199–206, 2008.

[32] C. Ziemkiewicz, R. J. Crouser, A. R. Yauilla, S. L. Su, W. Ribarsky, and R. Chang, "How locus of control influences compatibility with visualization style," *Proc. IEEE Conf. Visual Analytics Science and Technology*, Providence, RI, USA, pp. 81–90, 2011.

[33] S. D'Mello, A. Olney, C. Williams, and P. Hays, "Gaze tutor: A gaze-reactive intelligent tutoring system," *Int. J. Hum.-Comput. Stud.*, vol. 70, no. 5, pp. 377–398, May 2012.

[34] J. L. Sibert, M. Gokturk, and R. A. Lavine, "The reading assistant: eye gaze triggered auditory prompting for reading remediation," *Proc. 13th ACM Ann. Symp. User Interface Software and Technology*, pp. 101–107, 2000.

[35] F. Alt, A.S. Shirazi, A. Schmidt, and J. Mennenöh, "Increasing the user's attention on the web: using implicit interaction based on gaze behavior to tailor content," *Proc. 7th Nordic Conf. Human-Computer Interaction*, Copenhagen, Denmark, pp. 544–553, 2012.

[36] N. Jaques, C. Conati, J.M. Harley, and R. Azevedo, "Predicting Affect from Gaze Data during Interaction with an Intelligent Tutoring System," *Proc. 12th Int. Conf. Intelligent Tutoring Systems*, Honolulu, HI, USA, pp. 29–38, 2014.

[37] S. D'Mello, C. Mills, R. Bixler, and N. Bosch, "Zone out no more: Mitigating mind wandering during computerized reading," *Proc. 10th Int. Conf. Educational Data Mining*, Wuhan, China, pp. 8–15, 2017.

[38] S. Folker, H. Ritter, and L. Sichelschmidt, "Processing and Integrating Multimodal Material — The Influence of Color-Coding," *Proc. Annu. Meet. Cogn. Sci. Soc.*, vol. 27, no. 27, pp. 690–695, 2005.

[39] E. Ozcelik, I. Arslan-Ari, and K. Cagiltay, "Why does signaling enhance multimedia learning? Evidence from eye movements," *Comput. Hum. Behav.*, vol. 26, no. 1, pp. 110–117, Jan. 2010.

[40] S. Kalyuga, "Enhancing Instructional Efficiency of Interactive E-learning Environments: A Cognitive Load Perspective," *Educ. Psychol. Rev.*, vol. 19, no. 3, pp. 387–399, Aug. 2007.

[41] N. Moacdieh and N. Sarter, "Display Clutter: A Review of Definitions and Measurement Techniques," *Hum. Factors*, vol. 57, no. 1, pp. 61–100, Feb. 2015.

[42] G. Ellis and A. Dix, "A Taxonomy of Clutter Reduction for Information Visualisation," *IEEE Trans. Vis. Comput. Graph.*, vol. 13, no. 6, pp. 1216–1223, Nov. 2007.

[43] G. Carenini, C. Conati, E. Hoque, B. Steichen, D. Toker, and J. T. Enns, "Highlighting Interventions and User Differences: Informing Adaptive Information Visualization Support," *Proc. ACM SIGCHI Conf. Human Factors in Computing Systems*, Toronto, Canada, pp. 1835–1844, 2014.

[44] M.C. Dyson and M. Haselgrove, "The influence of reading speed and line length on the effectiveness of reading from screen," *Int. J. Hum.-Comput. Stud.*, vol. 54, no. 4, pp. 585–612, Apr. 2001.

[45] A.M. Lund, "Measuring Usability with the USE Questionnaire12," *Usability Interface*, vol. 8, no. 2, pp. 3–6, 2001.

[46] A. Kuznetsova, P.B. Brockhoff, and R.H.B. Christensen, "lmerTest package: tests in linear mixed effects models," *J. Stat. Softw.*, vol. 82, no. 13, 2017.

[47] Y. Benjamini and Y. Hochberg, "Controlling the false discovery rate: a practical and powerful approach to multiple testing," *J. R. Stat. Soc. Ser. B Methodol.*, vol. 57, no. 1, pp. 289–300, 1995.

[48] N. Kodagoda, B.L.W. Wong, C. Rooney, and N. Khan, "Interactive Visualization for Low Literacy Users: From Lessons Learnt to Design," *Proc. ACM SIGCHI Conf. Human Factors in Computing Systems*, New York, NY, USA, pp. 1159–1168, 2012.

**S. Lallé, PhD,** is a Research Associate at the University of British Columbia (Canada). He received his Ph.D. degrees in Computer Science from the University of Grenoble (France) in 2013, and was a graduate student visitor at Carnegie Mellon University (USA) in 2012. His areas of interest include User Modeling, User-Adaptive Systems, Intelligent Tutoring Systems, and Machine Learning. His peer-reviewed publications in these fields have won the best paper award at IVA'16 (Intelligent Virtual Agents) and have been nominated twice for the best student paper award (at AIED'13 and TICE'12). He is a member of the Board of Distinguished Reviewer of *ACM Trans. Interact. Intell. Syst.*

**D. Toker** is a PhD candidate at the University of British Columbia (Canada), and received his MSc degree there in 2013. His areas of interests include User Modeling, InfoVis, Eye-Tracking, and Machine Learning. His peer-reviewed publications in these fields have won the best paper award at UMAP'14 (User Modeling, Adaptation and Personalization) and have been nominated once for the best paper award (at IUI'14). He is finishing up his PhD and plans to enter industry as statistics consultant and machine learning engineer.

**C. Conati, PhD,** is a Professor of Computer Science at the University of British Columbia,(Canada). Her research goal is to integrate research in Artificial Intelligence (AI), Cognitive Science and Human Computer Interaction (HCI) to make complex interactive systems increasingly more effective and adaptive to the users' needs. Her areas of interest include User-Adaptive Interaction, User Modeling, Intelligent Tutoring Systems and Affective Computing. She has over 100 peer-reviewed publications in these fields, and her research has received awards from a variety of venues, including UMUAI (2002), IUI (2007), UMAP (2013, 2014) and TiiS (2014). Dr. Conati is an associate editor for UMUAI, ACM TiiS, IEEE Trans. Affect. Comp., and IJAIED. She has served as Program or Conference Chair for several international conferences including UMAP, IUI, and AI in Education.